\documentclass[twocolumn]{revtex4}

\usepackage{amsmath,amssymb,amsfonts}

\usepackage{algorithmic}
\usepackage{graphicx}
\usepackage{textcomp}
\usepackage[sc]{mathpazo}
\usepackage[T1]{fontenc}
\usepackage{microtype}
\usepackage[english]{babel}
\usepackage{ragged2e}
\usepackage{amsthm}
\newtheorem*{definition}{Definition}
\usepackage{natbib}

\usepackage{enumitem}
\setlist[itemize]{noitemsep}

\usepackage{hyperref}
\usepackage{subfiles}
\usepackage{graphicx}
\graphicspath{{./Images/}{../Images/}}

\def\BibTeX{{\rm B\kern-.05em{\sc i\kern-.025em b}\kern-.08em
T\kern-.1667em\lower.7ex\hbox{E}\kern-.125emX}}

\begin{document}

\title{Quantum Advantage on Proof of Work}

\author{Dan A. Bard }
\email{dab49@kent.ac.uk}
\author{Joseph J. Kearney }
\email{jjk30@kent.ac.uk}
\author{Carlos A. Perez-Delgado}
\email{c.perez@kent.ac.uk} 

\affiliation{School of Computing, University of Kent, Canterbury, Kent CT2 7NF United Kingdom}

\begin{abstract}
	Proof-of-Work (PoW) is a fundamental underlying technology behind most major blockchain cryptocurrencies. It has been previously pointed out that quantum devices provide a computational advantage in performing PoW in the context of Bitcoin. Here we make the case that this quantum advantage extends not only to all existing PoW mechanisms, but to any possible PoW as well. This has strong consequences regarding both quantum-based attacks on the integrity of the entirety of the blockchain, as well as more legitimate uses of quantum computation for the purpose of mining Bitcoin and other cryptocurrencies. For the first case, we estimate when these quantum attacks will become feasible, for various cryptocurrencies, and discuss the impact of such attacks. For the latter, we derive a precise formula to calculate the economic incentive for switching to quantum-based cryptocurrency miners. Using this formula, we analyze several test scenarios, and conclude that investing in quantum hardware for cryptocurrency mining has the potential to pay off immensely. 
\end{abstract}

\maketitle

\flushleft
\section{Introduction}\label{sec:intro}

Blockchain systems have become an integral part of modern financial society with their use reaching beyond the storage of value and cryptocurrencies into the wider financial market \cite{crosby2016blockchain}. One of the core tenets of these systems is that decisions about what data is immutably written to the blockchain's ledger, and therefore what is made a permanent entry on the chain's state going forwards, is made by consensus between nodes connected to and storing the ledger's information.

Although, consensus can be achieved utilizing several different methods \cite{8400278}, Proof of Work (PoW) powered blockchains currently account for more than 90\% of the current market share \cite{anand2016colored} and include some of the largest cryptocurrencies such as Bitcoin and Ethereum. These two blockchains alone account for a market cap of over US\$430 billion (approximate as of December 2020) \cite{CoinMarketCap}. This demonstrates that considerable financial assets are stored and maintained by blockchains, their transactions and therefore the underlying consensus algorithms.

In this paper we focus on the PoW mechanisms of blockchains. We show that quantum computers give a quadratic advantage in PoW efficiency; not just for all existing protocols but for any possible PoW protocol that relies on computational work being done.

Unlike many other cryptographic standards, blockchain systems intrinsically tie the protected asset (the ledger) with the encryption systems used. It has been previously shown that this makes blockchains particularly vulnerable to quantum attacks \cite{kearney, aggarwal2017quantum}. The main concern is that replacing the cryptographic protocols that build a blockchain with `post-quantum' ones is extremely more difficult than with more traditional cryptographic uses \cite{kearney, aggarwal2017quantum}. Several predicted timelines \cite{van2013blueprint,mosca} pin the year 2035 as when we can expect quantum computers to reliably be able to break current mainstream cryptographic protocols such as RSA2048 and ECDSA. These two key facts make these concerns timely and pressing.

Within most blockchain technologies, PoW underpins the protocols' consensus algorithm and because the consensus algorithm determines which transactions and actions performed on the network are integrated into the chain. This gives a quantum actor a potentially much stronger ability to control the decision-making in the blockchain.  From a cybersecurity perspective, when one actor (or group of actors) can reliably force all decisions in the blockchain, it is called a `51\%' attack \cite{ye2018analysis}.  In the first part of this paper, we will describe how quantum actors can much more reliably, and with much fewer resources than any classical counterpart, perform `51\%' attacks.

In the second part of this paper, we will consider a much less sinister, and much more profitable, use of quantum resources. Given the quadratic increase in PoW efficiency, one may consider using a quantum computer to \emph{`mine'} Bitcoin or some other cryptocurrency (\emph{mining} is the act of performing PoW in order to help the blockchain arrive at a consensus). Performing this task generally involves economic remuneration to the \emph{`miner'}. A quantum cryptocurrency miner can potentially require fewer clock cycles, a lot less energy, and dissipate a lot less heat, in order to mine the same amount of cryptocurrency as a classical computer counterpart. Whether this makes the endeavor profitable, of course, will depend on both the initial cost, and operating costs of such a quantum device. We explore these questions in Section \ref{sec:eqn}.

First, however, in the following section, we will discuss PoW as it is understood today, in more technical and formal detail. Then, in Section \ref{sec:grover}, we summarize the quantum technologies that can be deployed for PoW, and the advantages in doing so. Finally, we conclude with a summary of results, and a discussion on the future outlook for PoW. 

\section{Proof of Work}\label{sec:pow}

Consensus algorithms within blockchain technologies are critical to the running of the protocol and PoW is the most commonly utilized mechanism. It is used to ensure miners act honestly according to the rules of the blockchain protocol \cite{antonopoulos2014mastering}. It was adapted as a mechanism for consensus across a blockchain  by Satoshi Nakamoto \cite{nakamoto2019bitcoin,  back2002hashcash}.  PoW is widely used partly due to the utilization of Bitcoin's technologies and code base within a large amount of subsequent projects, but also because it is a highly secure mechanism for ensuring the good nature of mining nodes and because it lends itself well to distributed networks. 

Blockchain consensus employs the concept of the longest chain. The longest chain is typically the valid chain that a majority of the network holds as the state of the blockchain.  While a miner can create a malicious block and add it to the network trivially, it will not be accepted by a majority of the nodes, as other peers on the network will reject the block and choose an alternate proposed block, therefore, excluding the malicious block from the longest chain.  If a malicious user controls a majority of the network's computational power,  they could potentially overwhelm this consensus mechanism by adding blocks to the chain faster than the rest of the network can compete, meaning they consistently have the longest chain.  This means that the user could gain overall control of what is included into each block.  This is known as a `51\%' attack and is the most damaging threat to a blockchain's integrity. 

In PoW-based systems a user proposing a new block must perform a computationally-intensive task. The successful completion of this task must be easily verified by other users on the network.  Miners must expend a non-trivial amount of resources---usually computational \emph{work} and its associated costs, such as electricity and heat. This incurs a sunk cost for the miner that will be lost if the block they present is malicious or malformed.    

The Bitcoin PoW algorithm employs a NP-Complete problem where the goal is to create a hash digest based on a given input string \cite{ren2019analysis}. This hash digest is required to be in a specific form. This form is dictated by a target value (some integer in the range $[0,2^{256}]$) and Bitcoin miners must compute the hash digest that has an equal to or smaller value than the target. The target value is determined by the difficulty value of the blockchain network, which is altered depending on the computational power on the network as a whole as determined by the network's current \emph{hash-rate} (leading to the final target value being in the range of $[0,(2^{256} - difficulty)]$). Within Bitcoin, the difficulty value is changed according to the current computational power on the network once every 2016 blocks\cite{antonopoulos2014mastering} in order to maintain a block time of approximately 10 minutes. While this example is taken from Bitcoin, this is applicable to any network which utilizes PoW.

The hash is calculated using the block header, which is constant for a specific block, and a nonce, which is changed repeatedly by the miner, to create different hash digests in the hope of finding a digest that fits the requirements for the block. As noted earlier, this problem is NP-Complete. The best known classical algorithms for solving PoW scale exponentially to the size of the difficulty (which in turn is bounded by the size of the hash itself).

It is important to note that while hash-based PoW uses a NP-Complete problem, this does not necessarily have to be the case.  It \emph{must} be the case, however, that the miner expend a non-trivial amount of work, and that this expenditure can be verified by other users of the blockchain in a relatively trivial manner. In other words, let $TC_V$ be the time complexity for verification and $TC_S$ be the time complexity for the miner to solve the problem. Then, any PoW mechanism must guarantee that:

\begin{equation}\label{eq:hard}
	TC_V \ll TC_S.
\end{equation}

Clearly, any NP-Complete problem will satisfy the equation above. More generally, however, any PoW algorithm must satisfy the following requirements:

\begin{definition}[Proof of Work]\label{def:pow}A  computational problem can be considered as a PoW problem if it satisfies the following two requirements
	
	\begin{enumerate}
		\item The computational complexity of the problem must satisfy Eq.\ \ref{eq:hard},
		\item The difficulty of the problem must be easily \emph{tuneable} with a parameter.
	\end{enumerate}
\end{definition}

Requirement 1 has been explained above. Requirement 2 is an important requirement for the continued health of the blockchain network over time. As the computational power of miners increases, this parameter needs to be re-tuned to keep PoW as a meaningful deterrent against rogue miners.

In the following section we will explore the quantum computational advantage in PoW as described here. We will then explore the cybersecurity threat of quantum attacks on blockchain networks. Finally, we will analyze the possibility, and possible profit, of using this quantum advantage for the more benign purpose of more efficient cryptocurrency mining.

\section{Quantum Advantage for PoW}\label{sec:grover}

When discussing quantum advantage for computational tasks, two main types of algorithms are most often cited. The first is the subgroup-finding algorithms based on Shor's seminal work\cite{shor1994algorithms}. These types of algorithm provide a exponential advantage on problems including factoring and discrete logarithm. Though this is only a relatively small set of problems, it covers a large area of the cryptographic landscape. The other type are the quantum search algorithms based on Grover's algorithm\cite{grover1996fast,GroverOptical}. Whilst quantum search algorithms provide a more modest quadratic advantage over classical, their very broad applicability makes them extremely versatile, and central to our discussion.

The quantum search algorithm, as its name suggests, allows one to search \emph{any} (including unsorted and unstructured) data-set $S$, of cardinality $N = |S|$ for certain items that fulfill some condition, or is an element of some subset $C \subseteq S$. This condition is specified, in the quantum algorithm, as a black box or oracle $O$ that takes as input one register containing an element of $x \in S$, and an ancilla qubit, which is set to $1$ if $x \in C$ and $0$ otherwise. The importance of this algorithm is that it runs in \emph{total time} $O(\sqrt{N})$, and makes $O(\sqrt{N})$ queries to $O$. This oracle can be, and in practical uses often is, replaced by a quantum circuit or subroutine program $a$ that computes whether $x$ satisfies the required condition, or is an element, of the subset $C$.

In particular, one may consider a decision problem $D$ that is NP-Complete. Let $I$ be its input set, and $S \subseteq I$ the solution set. Given that the problem is in NP, there exists an efficient (polynomial-time) algorithm $a$ that can compute, on input $x$, whether $x \in S$. This in turn implies that a quantum search algorithm can solve $D$  in total time $O(\sqrt{N}) = O(\sqrt{2^n})$, where $n$ is the input size in bits. Because $D$ is NP-Hard, there is \emph{no} (known) classical algorithm that can solve $D$ in time substantially better than $O(2^n)$.

It should be now clear why the quantum search algorithm is of central importance to any discussion of quantum advantage for PoW. As discussed previously, most PoW systems today require the miner to find a SHA-x hash for a pre-determined string, that is under a certain value. This problem is NP-Complete.
Hence, a quantum computer with a memory register large enough to run Grover's algorithm on the necessary hash size, would be able to gain a quadratic  advantage over any classical device---including purpose-built ASICs.

To illustrate this, we can consider a toy example in which a classical brute-force search algorithm which runs in time precisely $2^n$, and a quantum search one that runs in precisely time $\sqrt{2^n}$. On input size $n=2$, the quantum algorithm is only twice as fast as the classical one. On input size $n=256$ the quantum algorithm will run $3.4 \times 10^{38}$ times faster. Compare this to ASIC chips that typically provide a speed-factor advantage of approximately $1 \times 10^{4}$.

We can also perform a more realistic analysis. Running a quantum search algorithm (assuming no error correction) on SHA-256 hashes requires roughly 512 qubits. Estimates by major quantum computer manufacturer predict such quantum computers will be available in 2023\cite{gambetta_2020}. At today's reported quantum computer clock-speeds\cite{arute_et_al._2019} (barring any major improvements) we can thus expect the equivalent of $4 \times 10^{7}$ calculations performed per second which, using Grover's algorithm, leads to the equivalent of $1.6 \times 10^{15}$ hashes computed per second (H/s).

\begin{figure}
	\includegraphics[width=.5\textwidth]{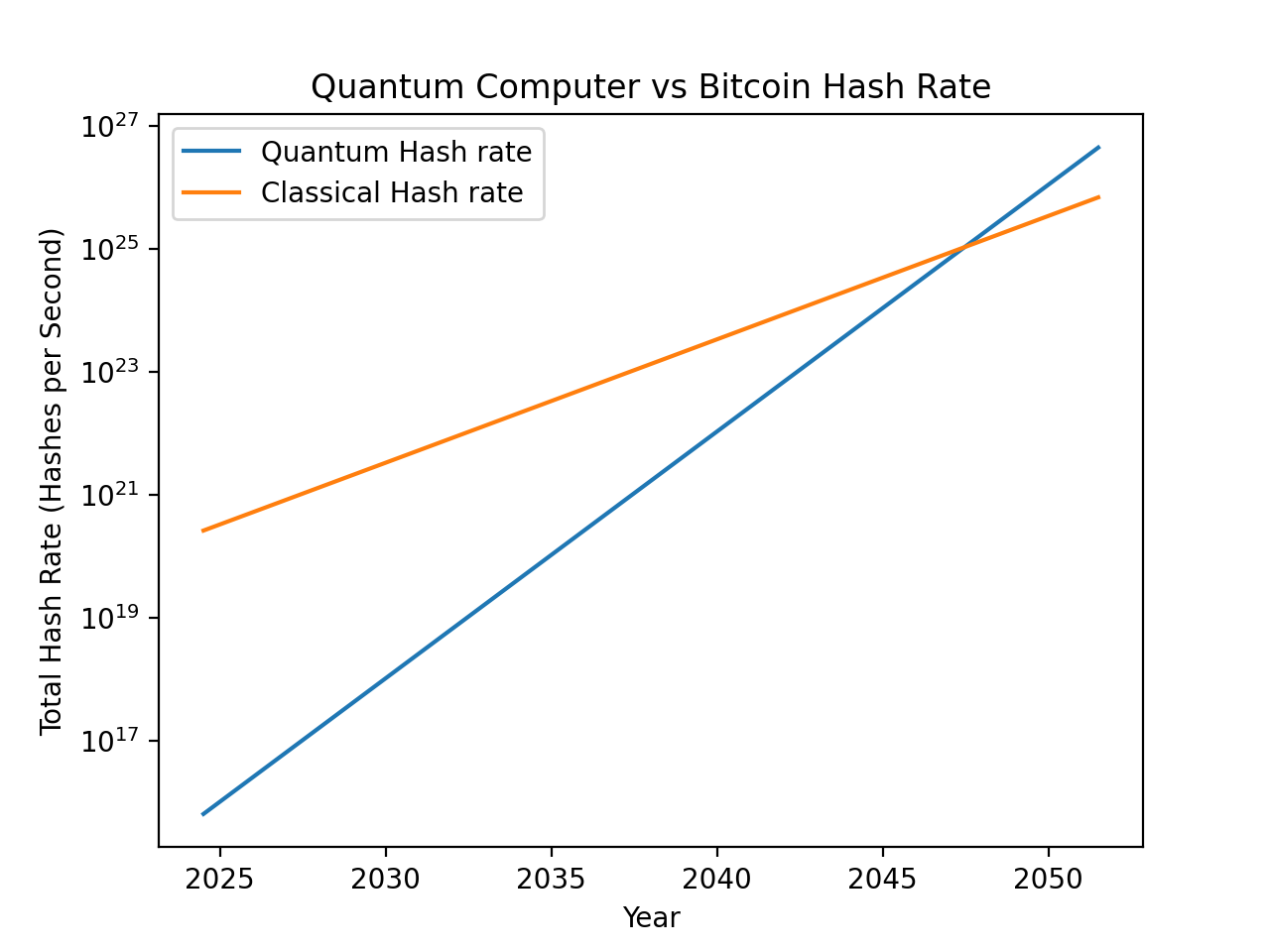}
	\caption{\textbf{Bitcoin network hash rate \emph{vs.} single quantum computer.} The graph shows the hash rate growth over time of the entirety of the Bitcoin network, compared to that of a \emph{single} quantum computer. Future data-points are extrapolated from current hash-rates, and assumes growth-rates for both quantum and classical technologies in line with current Moore's Law trends. See the main text for further details.}
	\label{fig:Graph1}
\end{figure}

Fig. \ref{fig:Graph1}  plots the Bitcoin network hash rate using the most current value of $130 \times 10^{18}H/s$\cite{BitcoinHashRate} against a quantum computing technology that starts at 40 MHz\cite{gambetta_2020}, and both increasing over time at the same rate, as dictated by Moore's Law. This gives an estimated timeframe of approximately 27 years until a \emph{single} quantum computer will be capable of completely out-mining the rest of the network, and hence be able to take over complete control of it (a successful $51\%$ attack). 

This prediction, however, is perhaps overly conservative for a couple of reasons. The first is that we consider the speed-increase in clock-rate for both quantum computers and classical computers to be the same. In reality, classical computers are known to be at the tail-end of Moore's Law's\cite{moore1965cramming} logistic-curve rate-of-growth\cite{shalf2020future}. Meanwhile, we can expect quantum computers, which are in their infancy, to over-achieve this rate-of-growth\cite{ghose_2019}.

Furthermore, this comparison has been made on the Bitcoin network, which has, by far, the largest hashing power of all  blockchains \cite{coinmetrics_2020}. Other comparatively smaller blockchain networks would be vulnerable far sooner than suggested here. For example, if network hash rates of blockchains such as Monero (1.28 Giga-Hashes per Second ($GH/s$))\cite{klemens_2020} or Ethereum Classic (6.43 Tera-Hashes per Second ($TH/s$))\cite{ethereumclassichashratechart} do not improve in the coming years, we could expect them to be vulnerable to a quantum 51\% attack as soon as there is a quantum computer with sufficient quantum memory, which is predicted to happen roughly in 2023\cite{gambetta_2020}.

In short, not only do quantum computers provide an \emph{asymptotic quadratic} efficiency increase for current PoW systems, they do so for any likely possible PoW system as well. Compare this to custom-built ASIC chips which also provide a speed increase in mining crypto-currencies, but are however limited to constant-factor speed-increases. This results in a single quantum computer being able to launch devastating attacks on the cryptocurrency network, in the foreseeable future.

Of course, this \emph{`single quantum computer'} attack would only work against a cryptocurrency network that is, at least for the most part, composed of classical miners. If a sizable portion of a cryptocurrency's miners were to move to quantum hardware, this would protect the entire network from quantum 51\% attacks. In the next section we explore legitimate  uses of quantum technology for PoW-based cryptocurrency mining. As we shall see, there may also be definite profit motives for individual cryptocurrency miners to invest in and adopt quantum technologies.

\section{The Profitability of Quantum Cryptocurrency Mining}\label{sec:eqn}

In previous sections we studied the cybersecurity threat posed by quantum-led 51\% attacks on blockchain networks. These attacks, while largely inevitable, are time-wise a bit far off---at least for the larger cryptocurrencies such as Bitcoin. The reason for this is that for a successful attack a quantum computer must have as much (or more) PoW computational power as the rest of the network combined.

Here, we will study the viability of using a quantum computer for the purpose of legitimately mining a cryptocurrency such as Bitcoin. In order to do \emph{this} effectively and profitably, a quantum computer doesn't have to be more powerful than the whole network, it only needs to be more efficient (in terms of resource-cost per block approved by the network) than a single classical miner. Hence, we can expect quantum supremacy, in the field of cryptocurrency mining, to be achieved much sooner than the previously discussed dates given for 51\% quantum attack viability.

We will first set out to derive a general equation that can be used to calculate the potential profitability of quantum-assisted cryptocurrency mining. We will then apply this equation to various credible scenarios, and give estimates of near-future profitability.

\subsection{Profitability Calculation}\label{sec:eqn_main}

In this section we will be setting out an equation to calculate whether mining on a classical or a quantum entity is more profitable. The primary element to be considered when making this calculation is the income from any device mining blocks on a blockchain. This is based on the probability of mining a block during the time it takes for a new block to be generated. This exact value varies among blockchains with a new block being generated on average every 600 seconds within Bitcoin \cite{project_2009} and approximately every 15 seconds within Ethereum \cite{io_2015}. However, this value can be generalized, since the relation between block generation and the probability of mining a specific block will be the same across all PoW based blockchains. This block time is controlled by the difficulty of a particular blockchain in relation to the hash size in bits defined by the blockchain's architecture \cite{garay2015bitcoin}. This is changed periodically in order to maintain a consistent block time and so, across a larger timescale, the time taken to generate a new block can be averaged dependent on blockchain. Based on these values and the given hash rate of any considered classical miner, we can say that the probability of mining a block is defined as:

\begin{equation}\label{eq:1}
	P_{C} =\frac{H_C  t}{\frac{\eta D}{t}}
\end{equation}

Where $P_C$ is the probability of mining a block from a classical device, $H_C$ is the hash rate of the classical device, $t$ is the block time, $D$ is the difficulty of the blockchain network and $\eta$ is the hash size.

The denominator of \ref{eq:1} is the calculation for the total network hash rate of any one network. This can then be simplified to:

\begin{equation}
	P_{C} =\frac{H_C t^{2}}{\eta D},
\end{equation}

as derived from the hash rate of any one device divided by the total network hash rate \cite{BitcoinHashRate}. As discussed in Sections \ref{sec:pow} and \ref{sec:grover}, as blockchain technologies utilize NP-Hard problems for PoW and as $D$ determines the complexity of said problems, $D$ is the value where the quadratic increase in efficiency can be applied. Due to this advantage the probability of mining a block on any quantum device based on a given equivalent hash rate can then be defined as:

\begin{equation}
	P_{Q} =\frac{H_Q  t^{2}}{\eta \sqrt{D}}
\end{equation}

Where $P_Q$ is the probability of mining a block from a quantum device and $H_Q$ is the equivalent hash rate of that device.

These probabilities, when taken across any given operational timespan can then be used to calculate the overall income across said timespan for any given blockchain, taking into account a conversion into fiat currency, defined as a function $f$. As the exact conversion between cryptocurrencies and real world fiat currency can vary, this has been abstracted to a single function. The exact reward gained per block mined is another element which varies based on the cryptocurrency being considered and duration of the operating period. When performing the profitability calculations, this needs to be taken into careful consideration as, for some cryptocurrencies, the block reward can change across the lifespan of any particular cryptocurrency. For example, Bitcoin halves its block reward every 210,000 blocks, meaning that though it originally rewarded 50 bitcoins (BTC) per block mined \cite{meynkhard2019fair}, the current value is 6.25 BTC. The reward is expected to  approach 0 by approximately 2140 \cite{controlled_supply}.

Taking these elements into account, the total income over the given timespan can be calculated as the following for classical miners:

\begin{equation}
	I_{C} = f\left(\frac{T}{t} \cdot P_{C} B\right),
\end{equation}

where $I_C$ is the income for a classical miner across the timespan $T$, and $B$ is the block reward for the considered blockchain.
The following holds for quantum miners:

\begin{equation}
	I_{Q} = f\left(\frac{T}{t} \cdot P_{Q} B\right),
\end{equation}

where $I_Q$ is the income for a quantum miner across $T$.

Following this, we can bring in the initial cost of any particular device in order to calculate the point at which the given device becomes profitable whilst operating on the network across $T$. Once this value becomes greater than 0, it is then deemed to be profitable to run the miner on the blockchain network. As discussed in Section \ref{sec:pow}, miners are required to expend energy (in the form of computation) to ensure honesty between parties. This is considered here as the operational costs of any given device.
From this, the profit returns for classical miners can be determined as:
\begin{equation}
	R_{C} = I_{C} - (T \cdot O_{C}) - S_{C},
\end{equation}
where $R_C$ is the profit, $O_C$ is the operating costs and $S_C$ is the setup costs for the classical device.
The profit calculation for quantum miners is as follows:
\begin{equation}
	R_{Q} = I_{Q} - (T \cdot O_{Q})-S_{Q},
\end{equation}
where $R_Q$ is the profit, $O_Q$ is the operating costs and $S_Q$ is the setup costs for the quantum device.

From these two equations, we can then calculate a profit ratio ($G$): 

\begin{equation}
	G = \frac{R_{C}}{R_{Q}}.\label{eq:golden}
\end{equation}

The above equation is particularly important: $G = 1$ is the inflection point at which quantum  and classical technologies are equally feasible. Values of $G$ less than 1 imply that  the quantum miner in question is more profitable than a classical one, even after factoring in initial investment costs considered in the calculation.

Eq.\ \ref{eq:golden} can be expanded, using the previous equations, to:

\begin{equation}\label{eq:final}
	G = \frac{f\left(T  \cdot  \frac{H_C  t}{\eta D} \cdot  B\right) - (T \cdot  O_{C})-S_{C}}{f\left(T  \cdot  \frac{H_Q  t}{\eta \sqrt{D}} \cdot B\right) - (T \cdot O_{Q})-S_{Q}}
\end{equation}

The above equation has many practical uses. For one, it allows one to \emph{`plug in'} various known values, like research and development and other initial investments necessary to jump-start a quantum crypto-currency mining operation, along with running costs for both classical and quantum mining, and decide whether the investment in quantum mining can pay off. It can also be used---as we do below---to estimate the timescales at which quantum cryptocurrency mining can become a profitable enterprise.

An important fact to emphasize is that Eq.\ \ref{eq:golden} takes into account the introduction of further quantum computing machines onto the network. This is because difficulty is defined at the protocol level of a blockchain as a mechanism to ensure that the block time stays within certain bounds. For example the Bitcoin blockchain's difficulty operates so that over a period of time, if the block time exceeds or is less than 10 minutes, the difficulty of the PoW problem is corrected to bring the block time back in line with the pre-defined desirable time. This means that the introduction of quantum computers onto the network will in fact decrease the block time as they have a quadratic advantage over their classical peers. The Bitcoin protocol will thereby increase the difficulty of the PoW algorithm. This is then taken into account within our equation. Introduction of quantum computers into the mining ecosystem could potentially cause a dramatic increase in the difficulty. This means that the equation presented here will take into account new quantum computers mining the network as their inclusion will factor into difficulty.

The above is important for various reasons, but of particular import is the \emph{first mover vs. second mover advantages.} Being a first mover, that is, being the first to enter a market (in this case with a quantum miner) has definite advantages and is of particular interest to entrepreneurs and investors. A common concern among potential investors is that of making a large investment, only to arrive \emph{late} to a market, potentially ruining return-on-investment prospects. As we shall discuss in the last section of this paper, quantum mining has the peculiar property that the more quantum mining \emph{`competitors'} one has, however, the more profitable it may become for one to do quantum mining.

\subsection{Scenarios and Forecasts}

Using the equation derived earlier, we can analyze some possible near-future scenarios. The general goal will be to determine the profitability of quantum-based cryptocurrency mining. The cryptocurrency which shall be used for this investigation will be Bitcoin as this is currently the blockchain with the highest comparative market value\cite{CoinMarketCap}. This shall be performed utilizing the denominator of Eq.\ \ref{eq:final} which can be formalized with the target as such:

\begin{equation}
	f\left(T  \cdot  \frac{H_Q  t}{\eta \sqrt{D}} \cdot B\right) - (T \cdot  O_{Q})-S_{Q} \leq 1
\end{equation}

For our case-analysis scenario, let us consider using a cloud quantum computing service. IBM\cite{ibmquantumexperience}, amongst others, have announced for-profit cloud-based quantum computing services. This is a natural scenario to consider since most quantum computation in the near future is likely to involve cloud-based services\cite{devitt2016performing,castelvecchi2017ibm}. This scenario has a composite advantage as well: it obviates the need for an initial investment, requiring instead only that the potential miner pay the rolling costs of renting quantum CPU time from the cloud provider. It will allow us, within this analysis, to set $S_Q = 0$.

Next, let us consider a time-frame. According to the roadmap set out by IBM, a quantum computer which can run a quantum search algorithm on Bitcoin's hashing function can be expected by roughly 2023\cite{gambetta_2020}. To be conservative, we consider 2025 to be an estimated `year zero' in which a quantum computer can run a quantum search on hash-based PoW and so 01/01/2025 shall be used whenever a given date is required.

For our case scenario we are focusing on Bitcoin. This sets some further variables in our equation. These are $t = 600s$, $\eta = 2^{32}$\cite{antonopoulos2014mastering,nakamoto2019bitcoin} and $B = 3.125 BTC$\cite{meynkhard2019fair,buybitcoinworldwide}. As an additional part of the blockchain architecture, the difficulty is calculated and adjusted every 210,000 blocks in order for the block period to remain relatively constant. To provide a difficulty for this scenario, we plotted the historical difficulties and then the appropriate difficulty was extrapolated to our given date using polynomial curve of best fit. This provided a difficulty of $D = 4.2903 \times 10^{18}$. Though there are varying opinions of the future of Bitcoin difficulty\cite{kraft2016difficulty}, this matches the current trends.

The value of Bitcoin has had a general increasing trend year-on-year, however due to the volatile nature of cryptocurrencies, no single prediction can be made. Therefore the values shown in table \ref{tab:table1} will account for various BTC to USD conversion rates including the current price (as of 17/12/2020 this is \$23,536.12)\cite{CoinMarketCap}, the average price over the last 12 months (taken as the average closing price from 01/01/2020 until 17/12/2020, \$10,385.49), a predicted conservative price (\$31,000) and a predicted high-end price (\$100,000).

The final element of the equation to be assigned is the hash rate equivalent of the quantum computer. How the hashing power  will increase as the development of quantum computers continues is not known. Thus, we consider two possibilities.
In the first scenario, we take the clock-speed of (one of) Google's current quantum computers of $H_Q = 40 MHz/s$\cite{arute_et_al._2019}, and keep that value constant throughout time.
In the second, more plausible, scenario we increase the quantum computer's clock-speed according to Moore's Law. After four doubling cycles, we arrive at a clock-speed of $H_Q = 640 MHz/s$.

Table \ref{tab:table1} collects the calculations made for the various scenarios.

\begin{table}[]
	\begin{tabular}{|r|r|p{60pt}|}
		\hline
		$H_Q (MHz/s)$          & $f (USD)$        & $O_Q$    \\ \hline
		40                    & 23,536.12        & 6,258.27                     \\ \hline
		40                    & 10,385.49        & 2,761.51                     \\ \hline
		40                    & 31,000.00        & 8,242.92                     \\ \hline
		40                    & 100,000.00       & 26,590.06                    \\ \hline
		640                   & 23,536.12        & 100,132.28                   \\ \hline
		640                   & 10,385.49        & 44,184.12                    \\ \hline
		640                   & 31,000.00        & 131,886.68                   \\ \hline
		640                   & 100,000.00       & 425,440.90                   \\ \hline
	\end{tabular}
	\caption{This table shows the income generated over the period of a year in USD related to specified quantum computer clock speed and fiat currency conversions. In the third column $O_Q$ is calculated with $I_Q = 1 (USD)$}
	\label{tab:table1}
\end{table}

From these results, the best case scenario can be found when $H_Q = 640MHz/s$ and the market conversion result is $f = \$100,000$. In this case, as long as the operational cost of the quantum device (\emph{i.e.} the quantum cloud CPU time charged by the provider) is below $O_Q = \$425,440.90$ a year, a quantum miner would still be able to turn a profit.

\subsection{The Effects of Introducing Quantum PoW Technology}\label{sec:eqn_loop}

\begin{figure}[ht]
	\begin{center}
	\includegraphics[width=0.45\textwidth]{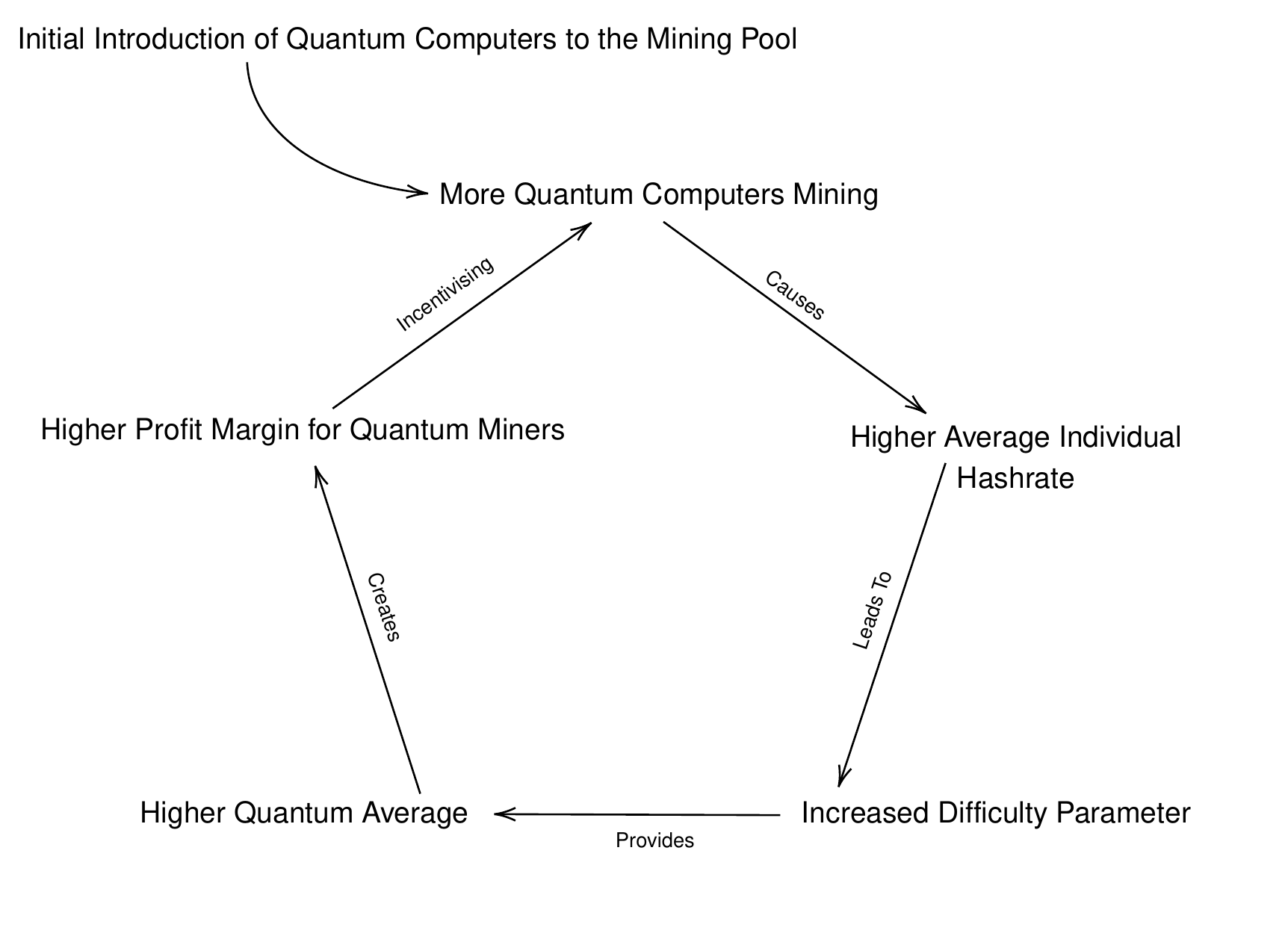}
	\end{center}
	\caption{\textbf{Self-propagating cycle of increasing quantum advantage on PoW networks.} Adding quantum miners to the cryptocurrency network increases the network's hash-rate. An increased hash-rate will raise the difficulty parameter. An increased difficulty parameter increases the relative quantum-advantage. This, in turn, increases the profitability of quantum-mining, which in turn motivates the introduction of more quantum miners.}
	\label{fig:graph2}
\end{figure}

Finally, presented in Figure \ref{fig:graph2} is a cascading \emph{virtuous cycle} that will propagate upon the introduction of quantum computers to a PoW based blockchain network. This will happen as they become profitable when compared to classical alternatives, according to Equation \ref{eq:final}.  Firstly, introducing quantum computers into a PoW based blockchain, as discussed, will consequently increase the hash-rate of the entire network, thereby shortening the average time it takes for the network to calculate a block.  According to a blockchains protocol this will cause an increase in the PoW difficulty parameter in order to recalibrate the block-time to the prescribed value.  

Increasing the difficulty parameter of PoW has been shown to solidify the quadratic advantage of quantum computers as miners. This advantage means that there will be greater incentive for investment into quantum mining technologies as the profit margin when compared with their classical counterparts will increase. This greater incentive will once again increase the number of quantum miners on the network, thereby decreasing the block-time and increasing the PoW difficulty in turn. This creates cycle within which quantum computing technologies will, eventually, completely replace classical miners, as the later cease to be cost-effective.

This cascading effect also has a security benefit for the network itself. As soon as the majority (roughly) of the miners are quantum, the network itself become impervious to 51\% attacks based on quantum advantage \emph{alone}. It would still be technically possible to mount such an attack, but such an attack would only succeed by using \emph{other} methods such as miner-collusion, rather than by merely leveraging quantum advantage.

Over time the increased difficulty parameter of PoW will lead to classical miners being made obsolete. The increase in difficulty will cause the PoW problem to become exponentially harder for both classical and quantum devices. However, the impact to classical miners is quadratically worse, over time, than the impact to quantum miners. Eventually, this will lead to all quantum miners being more cost-effective than classical miners (regardless of their initial setup costs).

\section{Conclusion}\label{sec:conc}

In closing, quantum computation gives a definite advantage over classical computation for the purpose of calculating PoW for blockchains. As we have seen, in Sec.\ \ref{sec:grover}, this quantum advantage can be used by an adversarial party, in order to attempt what are called 51\% attacks, on the cryptocurrency. The possibility of these types of attacks is, however, in the reasonably distant future. 

On the other hand, it is very unlikely that there ever will be a quantum-secure---or \emph{post-quantum}---alternative to hashing for the purposes of PoW. Not only is hashing-based PoW susceptible to quantum advantage, but so are other well-known PoW systems such as Zcash's use of the Birthday Paradox-based computational problem\cite{hopwood2016zcash}.

Moreover, it is unlikely that any PoW system can be derived is not susceptible to some form of quantum advantage. This is because PoW, by definition, requires a problem whose solution is hard to \emph{compute} (to ensure miners are required to do meaningful \emph{work} for their PoW), while being fairly easy to \emph{verify} (to ensure any third party can verify that the work has been performed). And these are exactly the type of problems where quantum search algorithms provide definite advantage over classical. 

This means that once a quantum computer \emph{does} exist that can attack the network in this way, there will be very little that can be done to safeguard the blockchain network against said attacks.

One possible avenue is to drop the use of PoW by the blockchain completely, and move to another consensus mechanism entirely---such as \emph{Proof of Space}\cite{dziembowski2015proofs}, or other alternatives\cite{bentov2014proof,alwen2017scrypt}.

Another safeguarding mechanism would be to move the entire cryptocurrency from ASIC miners to quantum miners. In Sec.\ \ref{sec:eqn}, we discussed the possibility of doing this. We showed that mining cryptocurrency, using quantum computation, can quickly become a profitable proposition. In Sec.\ \ref{sec:eqn_main} we gave a precise formula that allows one to calculate a potential profit of using quantum computation for PoW.

How profitable this is will clearly depend on considerations such as the running cost of a quantum computer, and the initial costs of setting one up. This latter cost can be removed if one chooses to use cloud quantum computation. We calculated the precise revenue that one can expect from mining bitcoin in 2025 across the period of a year using predicted available cloud quantum computing at that time to be between \$44,184.12 and \$425,440.90, depending on whether the most conservative or optimistic parameters are used. This variable is based on the conversion rate of Bitcoin and the exact hashing power of the quantum device at the time. Whether this is profitable will depend on how much quantum cloud CPU time is charged for at that time. The existence of secure remote quantum computing protocols such as \emph{blind quantum computation}\cite{barz2012demonstration}, means that a client can safely use a cloud quantum server for the purposes of mining Bitcoin, or other cryptocurrencies, without any interference from the server. In short, this shows a very likely profitable use of quantum computational resources in the coming decades. As shown in Fig. \ref{fig:graph2}, and described more generally in Sec.\ \ref{sec:eqn_loop}, the introduction of quantum computers into a mining ecosystem will make subsequent use of quantum computers even more profitable, as compared to classical computers---which in turn will become less profitable over time.

In closing, we've introduced the mathematical machinery necessary to understand, accurately, the impact of introducing quantum PoW technology into cryptocurrency ecosystems---used both by malicious, and non-malicious actors. A clear next step is to branch out the analysis we have done here to other blockchain consensus mechanisms that were outside the scope of this work.

Another clear next step is to take the work we have developed here, as well as real-world economic data, and use both together to create accurate, predictive, models. Several useful predictive models could be developed to help inform, say, investment strategies into quantum technologies, hedging strategies for cryptocurrency investors and miners, etc. We have made some simple predictions here, in Sec.\ \ref{sec:eqn_loop}. These simple models are meant, mostly, to showcase the power of the mathematical machinery we have introduced---and to hopefully motivate their use in creating more accurate, powerful, predictive models. Even our very simplistic models already suggest a trend however: we expect all PoW-based cryptocurrency mining to move to quantum platforms in the coming decades.


\begin{thebibliography}{10}
\bibitem{controlled_supply}
en.bitcoin.it/wiki/, ``Controlled supply.''
\end{thebibliography}
\bibliographystyle{ieeetr}

\end{document}